# DECOMPOSITION OF THE BRAZIL'S 5-YEAR DI FUTURES IN BASIS POINTS

An integrated framework combining Focus expectations, macro surprise, and sovereign risk (domestic vs. global CDS)

Gabriel de Macedo Santos

Instituto de Tecnologia e Liderança

**ABSTRACT**

This article proposes and implements an empirical, replicable, and interpretable framework to decompose, in basis points (bps), the daily variations of Brazil's 5-year DI Futures (DI5Y). The strategy combines three blocks: (i) macroeconomic and fiscal expectations extracted from the Central Bank of Brazil's Focus Report, transformed into daily changes; (ii) a supervised macro factor built using Partial Least Squares (PLS), which synthesizes changes in expectations and a high-frequency "surprise" measure; and (iii) a decomposition of sovereign risk via Credit Default Swaps (CDS) into global and domestic components, obtained by regressing CDS on external financial conditions variables (DXY, CRB, VIX, and UST10). The final step converts the factors into daily contributions in bps through a linear regression of $\Delta DI5Y$ on the three factors, enabling a cumulative historical decomposition that "adds up" with a constant and a residual. In the sample available in the final output (2015-01-13 to 2025-12-12; 2,741 observations), the model explains approximately 22.45% of the daily variance of $\Delta DI5Y$. The explained portion is dominated by domestic risk (CDS_dom), with a secondary but statistically relevant role for the macro factor. The residual remains high, pointing to limitations inherent to linearity and to the absence of additional drivers (monetary policy in event windows, term premium dynamics, liquidity, and positioning). Even so, the framework provides a clear quantitative narrative: how much of the daily (and cumulative) movement in DI5Y can be attributed to macro/central bank forces, Brazil risk, and external risk.

# 1. INTRODUCTION

Movements in the yield curve of emerging economies are, by construction, a hybrid object: they reflect expectations about inflation, economic activity, and monetary policy, but they also embed risk premia that are sensitive to fiscal shocks, political uncertainty, global liquidity conditions, risk aversion, and unobservable factors. In practice, interpreting "what moved the curve" is often based on qualitative narratives, with limited quantification that is comparable over time. This article has a straightforward objective: to build a simple and auditable procedure capable of quantifying, in basis points (bps), the daily impact of three blocks of drivers on Brazil's 5-year DI Futures (DI5Y): (i) macro/central bank; (ii) domestic risk; and (iii) external risk.

The starting point is the term-structure decomposition and factor-model tradition, which seeks to map yield dynamics into a reduced set of interpretable components (for example, through latent and macroeconomic factors, and through pricing regressions). In particular, regression- and synthetic-factor-based approaches provide a pragmatic path for translating high-frequency information into bps contributions (Adrian, Crump, & Moench, 2016). At the same time, for emerging markets, sovereign risk measures (such as CDS) respond quickly both to global shocks and to domestic idiosyncrasies, and decomposing them into "global vs. domestic" components is useful for economic attribution (Fender, Hayo, & Neuenkirch, 2012; Rodríguez, Dandapani, & Lawrence, 2019).

This paper's contribution is operational: to combine, in a single pipeline, (a) expectations from the Focus Report; (b) a supervised macro factor built via PLS; (c) sovereign risk decomposed into global and domestic components; and (d) a decomposition regression that converts factors into bps and produces an accumulated accounting of DI5Y movements. The result is an analytical artifact that can be put "into production" (with re-estimation) and used for historical interpretation and for ongoing monitoring that can be updated weekly and visualized on business days.

# 2. DATA AND PREPARATION

## 2.1. Focus Expectations (BCB)

Expectations are collected automatically via the Central Bank of Brazil (BCB) Application Programming Interface (API), using the annual market expectations endpoint. For each collection date, we obtain the median annual expectations for inflation (IPCA), the policy rate (Selic), economic activity (GDP), and fiscal variables (primary balance and nominal balance). The final expectations dataset is built in a "Date × Reference year" panel format, and is later reshaped into a set of horizon-specific columns, in the style of:

| Data | IPCA year | IPCA year_1 | IPCA year_2 | IPCA year_3 | Selic year | Selic year_1 | Selic year_2 | Selic year_3 | PIB year | PIB year_1 | PIB year_2 | PIB year_3 | Primario year | Primario year_1 | Primario year_2 | Primario year_3 | Nominal year | Nominal year_1 | Nominal year_2 | Nominal year_3 |
|---|---|---|---|---|---|---|---|---|---|---|---|---|---|---|---|---|---|---|---|---|
| 02/01/2004 | 6.00 | 5.00 | 4.50 | 4.00 | 13.85 | 13.00 | 12.00 | 11.50 | 3.66 | 3.66 | 3.66 | 3.66 | 4.25 | 4.25 | 4.00 | 3.75 | -3.00 | -2.35 | -2.50 | -2.20 |
| 05/01/2004 | 6.00 | 5.00 | 4.50 | 4.00 | 13.85 | 13.00 | 12.00 | 11.50 | 3.72 | 3.72 | 3.72 | 3.72 | 4.25 | 4.23 | 4.00 | 3.75 | -3.00 | -2.50 | -2.50 | -2.50 |
| 06/01/2004 | 6.00 | 5.00 | 4.50 | 4.00 | 13.85 | 13.00 | 12.00 | 11.50 | 3.66 | 3.66 | 3.66 | 3.66 | 4.25 | 4.25 | 4.00 | 3.75 | -3.00 | -2.45 | -2.50 | -2.15 |
| 07/01/2004 | 6.00 | 5.00 | 4.50 | 4.00 | 13.85 | 13.00 | 12.00 | 11.50 | 3.60 | 3.60 | 3.60 | 3.60 | 4.25 | 4.25 | 4.00 | 3.75 | -3.00 | -2.30 | -2.30 | -2.13 |
| 18/12/2025 | 4.35 | 4.09 | 3.80 | 3.50 | 15.00 | 12.00 | 10.50 | 9.50 | 1.80 | 1.80 | 1.80 | 1.80 | -0.50 | -0.60 | -0.40 | -0.12 | -8.43 | -8.66 | -7.84 | -7.20 |
| 19/12/2025 | 4.33 | 4.06 | 3.80 | 3.50 | 15.00 | 12.25 | 10.50 | 9.75 | 1.80 | 1.80 | 1.80 | 1.80 | -0.50 | -0.60 | -0.34 | -0.16 | -8.43 | -8.70 | -7.85 | -7.00 |
| 22/12/2025 | 4.33 | 4.06 | 3.80 | 3.50 | 15.00 | 12.25 | 10.50 | 10.00 | 1.80 | 1.80 | 1.80 | 1.80 | -0.50 | -0.59 | -0.35 | -0.18 | -8.43 | -8.61 | -7.90 | -7.00 |
| 23/12/2025 | 4.33 | 4.06 | 3.80 | 3.50 | 15.00 | 12.25 | 10.50 | 10.00 | 1.80 | 1.80 | 1.80 | 1.80 | -0.50 | -0.59 | -0.35 | -0.18 | -8.43 | -8.61 | -7.90 | -7.00 |
| 24/12/2025 | 4.33 | 4.06 | 3.80 | 3.50 | 15.00 | 12.25 | 10.50 | 10.00 | 1.80 | 1.80 | 1.80 | 1.80 | -0.50 | -0.59 | -0.35 | -0.18 | -8.47 | -8.61 | -7.90 | -7.00 |
| 26/12/2025 | 4.32 | 4.05 | 3.80 | 3.50 | 15.00 | 12.25 | 10.50 | 9.75 | 1.80 | 1.80 | 1.80 | 1.80 | -0.50 | -0.56 | -0.35 | -0.18 | -8.43 | -8.61 | -7.90 | -7.00 |

*Figure 1 - Data Structure of the Focus Expectations*

**Variable year:** Represents the expectation for the year corresponding to the survey date.

**Variable year_1:** Represents the expectation for the year following the survey date.

**Variable year_2:** Represents the expectation for the year two years after the survey date.

**Variable year_3:** Represents the expectation for the year three years after the survey date.

The sample retrieved via the BCB API runs from 2004 to 2025-12-26, corresponding to the last Focus Report release in 2025. The study focuses on the daily variation of expectations captured by the Focus Report over time, which is later used as input to the PLS regression.

### 2.2. DI5Y and the surprise measure

The 5-year DI Futures (DI5Y) series is also obtained through an automated data-collection process; however, in this case it is based on web scraping from the B3 (Brazil's stock exchange) website. The routine captures the end-of-day value for each trading session, building a daily time series consistent with the frequency of the other variables in the study. Since the central goal of this work is to attribute yield-curve movements to different blocks of drivers in a directly interpretable unit, the article explicitly adopts the daily change in DI5Y in basis points (bps) as the dependent variable. Operationally, the model's target is to decompose the daily change in DI5Y—i.e., to identify how much of the observed shift on each day can be attributed to the macro factor, domestic risk, and external risk—while preserving a bps-consistent accounting over time.

$$(1)\ \Delta DI5Y_t = dDI5Y_t^{bps}$$

To construct the supervised macro factor, we incorporate a high-frequency measure of economic surprise, because changes in Focus expectations alone may react with a lag to new information and relevant macro events. For this purpose, we use Bloomberg's "BCMPBRGR Index," which summarizes the "surprise" component associated with economic releases relative to market expectations. The treatment applied is deliberately simple and compatible with daily frequency: we transform the index into its daily change (diff), isolating the incremental information component that reaches the market from one day to the next. This series is used exclusively in the PLS estimation step for the macro factor, serving as an additional channel to capture contemporaneous information shocks that may affect DI5Y before they are fully reflected in the consolidated Focus expectations.

### 2.3. CDS

Decomposing sovereign risk requires a CDS time series for Brazil as well as global proxies. The transformations follow a well-established convention in empirical finance: log returns are used for price/index variables, while simple differences are used for rates and yields. Definitions:

$$(2)\ r(\log)_{x,t} = \ln(x_t) - \ln(x_{t-1})$$

$$(3)\ \Delta z_t = z_t - z_{t-1}$$

In the pipeline, CDS and variables such as DXY, CRB, and VIX enter as log returns; UST10 enters as a simple difference.

## 3. Methodology

### 3.1. A supervised Macro/Central Bank factor (PLS)

The first block aims to summarize, into a single time series, the information contained in multiple Focus expectations (across several horizons) and in a daily surprise measure. Instead of applying PCA (which is purely unsupervised), we use Partial Least Squares (PLS), which builds linear combinations of the regressors that maximize their covariance with the target ($\Delta DI5Y$). Conceptually, PLS answers the question: "Which combination of changes in expectations and macro surprise moves most systematically with $\Delta DI5Y$?"

Let $X_t$ be the vector of regressors (changes in expectations and the transformed surprise measure) and $y_t = \Delta DI5Y_t$. PLS extracts a score $f_t$ (one component) such that:

$$(4) \quad f_t = w'X_t \text{ with } w \text{ choosed for maximize } \text{Cov}(f_t, y_t)$$

In the pipeline, $X_t$ is standardized (StandardScaler) before PLS, and the final factor is normalized to have zero mean and unit standard deviation within the macro-block estimation window. Importantly, the factor's sign is anchored to ensure interpretability:

$$(5) \quad if\ Corr(f_t, y_t) < 0, so\ f_t := -f_t$$

After anchoring, $f_t$ is standardized. In the factor file, this series is saved as "Macro_Factor_PLS", and should be interpreted as a synthesized "macro/central bank shock": positive values correspond, by construction, to days in which the combination of changes in expectations and surprise tends to point to an increase in DI5Y. In addition, it is important to keep in mind that this factor primarily reflects economic expectations; even with the surprise indicator, it is not able to fully capture fluctuations driven by specific relevant data releases as they occur.

A crucial methodological note: because the factor is supervised using $y$ itself ($\Delta DI5Y$), this step is not a purely exogenous extraction. Thus, PLS should be understood as an information-compression (dimensionality-reduction) mechanism guided by $y$, rather than as a causal instrument. This does not invalidate the accounting decomposition in bps, but it does require caution when interpreting the statistical significance of the coefficient associated with the macro factor in the final step (see Section 5).

### 3.2. Decomposing CDS into global and domestic components

CDS reflects a combination of domestic and external factors that ultimately generate the country risk premium. To understand how internal and external forces directly affect the yield curve, it is necessary to decompose CDS into these components. Therefore, the second block decomposes CDS movements into a portion explained by global conditions (**"CDS_glob"**) and a residual interpreted as domestic risk (**"CDS_dom"**). The regression is estimated in returns/variations:

$$(6)\ CDS_t = \alpha + \gamma_1 DXY_t + \gamma_2 CRB_t + \gamma_3 VIX_t + \gamma_4 UST10_t + u_t$$

We then define:

$$(7)\ CDS\_glob_t = \widehat{CDS}_t\ (fitted\ values)$$

$$(8)\ CDS\_dom_t = \widehat{u_t}\ (residuo)$$

The intuition is straightforward: CDS movements that track variations in a stronger dollar (DXY), shifts in commodities (CRB), changes in risk aversion (VIX), and moves in long-term U.S. rates (UST10) are classified as "global"; the remainder is classified as "domestic." For emerging markets, this separation is often useful to distinguish external stress from local idiosyncrasies (Fender, Hayo, & Neuenkirch, 2012).

### 3.3. Decomposition regression of ΔDI5Y in bps

With the three factors in hand— MacroBC, CDS_dom, and CDS_glob —the final step translates these drivers into daily contributions in bps through a linear regression:

$$(9)\ \Delta DI5Y_t = \beta_0 + \beta_M MacroBC_t + \beta_D CDS_{dom,t} + \beta_G CDS_{glob,t} + \epsilon_t$$

The daily decomposition in bps is then:

$$(10)\ Macro\_Contribution_t = \beta_M MacroBC_t$$

$$(11)\ RiscoBR\_Contribution_t = \beta_D CDS_{dom,t}$$

$$(12)\ Global\_Contribution_t = \beta_G CDS_{glob,t}$$

$$(13)\ Residual_t = \Delta DI5Y_t - (\beta_0 + Macro\_Contribution_t + RiscoBR\_Contribution_t + Global\_Contribution_t)$$

Finally, for historical analysis, the contributions are accumulated:

$$(14)\ DI5Y\_change_{cumu,t} = \sum_{s \leq t} \Delta DI5Y_s$$

$$(15)\ MacroBR_{cumu,t} = \sum_{s \leq t} Macro\_Contribuition_s$$

$$(16)\ RiscoBR_{cumu,t} = \sum_{s \leq t} RiscoBR\_Contribuition_s$$

$$(17)\ Global_{cumu,t} = \sum_{s \leq t} Global\_Contribuition_s$$

$$(18)\ Residual_{cumu,t} = \sum_{s \leq t} Residual_s$$

$$(19)\ Const_{cumu,t} = \sum_{s \leq t} \beta_0 = \beta_0 t$$

The result is an accounting identity that should add up in cumulative terms (up to numerical error):

## 4. RESULTS

### 4.1. Effective sample and data availability

Although the macro factor is available starting in 2004, the full decomposition requires CDS_dom and CDS_glob, whose availability begins on 2015-01-13. The reason for choosing this start date is to observe the behavior of the variation in a more recent regime, yet one characterized by de-anchored expectations. Since the final step works on the intersection of dates (inner join), the sample in the final output is:

**Period:** 2015-01-13 to 2025-12-12

**Number of observations:** 2,741

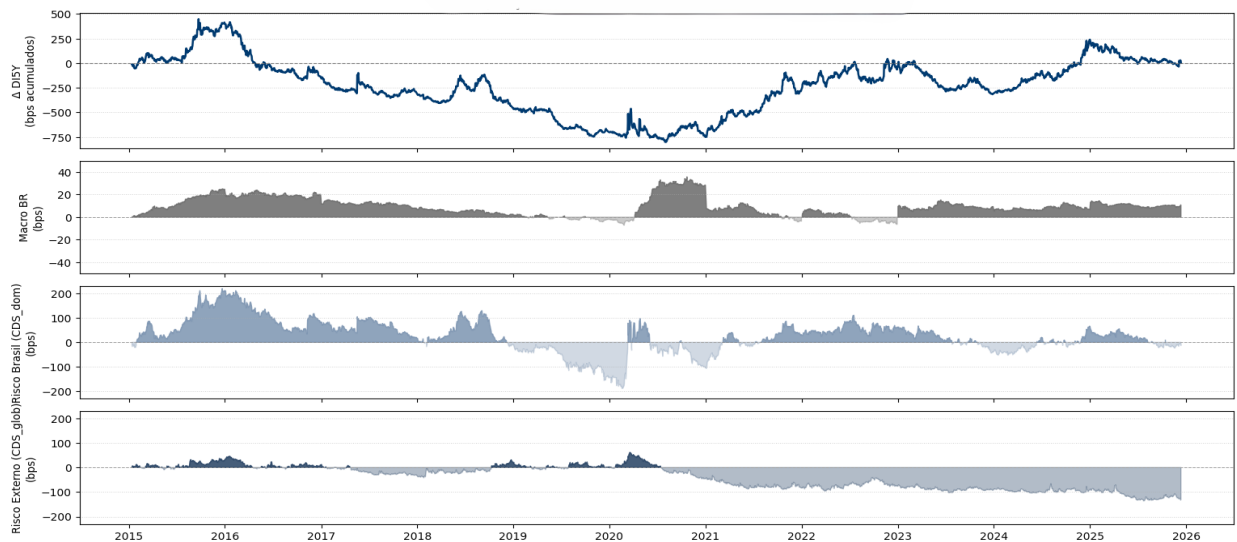

*Figure 2 - Decomposition of the 5Y DI Futures: Macro, Brazil Risk, External Risk*

This restriction is substantive: the decomposition covers a decade marked by episodes of both domestic and global stress, including major external shocks, domestic political crises, and domestic monetary policy cycles.

### 4.2. Baseline estimation of the decomposition regression

The regression in (9), estimated by OLS on the final-output sample, yields:

| Metric / Variable | Symbol | Value | p-value | Significance |
|---|---|---|---|---|
| Constant (intercept) | β₀ | 0,051434 | 0,8393 | Not significant |
| Macro Factor | β_M | 0,635428 | 0,0023 | Significant |
| Domestic CDS | β_D | 339,045,202 | ≈ 1,07×10^-130 | Highly Significant |
| Global CDS | β_G | 325,577,999 | ≈ 5,18×10^-29 | Highly Significant |
| R² | — | 0,224537 | — | — |
| Adjusted R² | — | 0,223687 | — | — |
| N (observations) | — | 3 | — | — |

*Table 1 – Output generated by regression (9).*

Immediate economic reading: both the domestic and the global CDS components have a very strong statistical association with daily ΔDI5Y; the supervised macro factor is also relevant, but with a smaller impact. The constant, while it enters the cumulative accounting, does not have a robust economic interpretation as a statistically non-zero "drift" at the daily horizon.

### 4.3. Economic magnitude: volatility and decomposition of the explained component

To compare magnitudes on a common metric, it is useful to look at the daily standard deviation (in bps) of ΔDI5Y and of the contributions:

| Series | Standard Deviation (bps) |
|---|---|
| Std(ΔDI5Y) | 150,631 |
| Std(Macro_Contribution) | 0,7732 |
| Std(RiscoBR_Contribution) | 65,172 |
| Std(Global_Contribution) | 28,679 |
| Std(Residual) | 132,646 |

*Table 2 – Daily standard deviation in bps.*

The standard deviation of the fitted component is approximately **7.1377 bps**, versus **15.0631 bps** for ΔDI5Y. This is consistent with **R² ≈ 0.2245**: the model captures a relevant share of daily variation, but the residual remains dominant.

A useful empirical feature is that the Macro, Brazil Risk, and Global contributions are nearly uncorrelated with each other in the sample (correlations close to zero). This allows, as a good approximation, an interpretation of the composition of explained variance. Considering the variance of the fitted component excluding the constant, the following approximate shares are obtained:

| Variable | Explained variance |
|---|---|
| MacroBR | ~1,17% |
| Risco Brasil | ~83,37% |
| Risco Externo | ~16,14% |

*Table 3 – Explained variance by variable.*

Qualitatively, what the model "explains" in ΔDI5Y is mostly domestic sovereign risk; external risk complements it; and the supervised macro factor acts as a fine-tuning adjustment and a directional signal in specific episodes.

**4.4. Cumulative decomposition: episodes and regimes**

The analytical gain of the framework becomes clear when contributions are accumulated. Three illustrative points (taken directly from the output) are:

**(i) Positive peak in cumulative DI — 2015-09-23**

**DI5Y_change_cumu = +449.0 bps**

Cumulative decomposition at this point:

| Accumulated Component | Value (bps) |
|---|---|
| Const_cum | 90,010 |
| Macro_cum | 186,477 |
| RiscoBR_cum | 1,894,902 |
| Global_cum | 196,635 |
| Residual_cum | 2,121,975 |

*Table 4 – Cumulative decomposition up to 09/23/2015.*

**Interpretation:** On that day, despite the Central Bank's attempts to intervene via the FX market, the U.S. dollar against the Brazilian real reached—up to that point—the highest nominal level in the series. In parallel, the deterioration in the political environment—partially captured by the Brazil Risk component (CDS_dom)—raised risk perceptions, reflecting the lack of consensus around a credible fiscal solution. The episode therefore combines a sizable shift attributed to domestic risk with an even larger residual share. This suggests that, while CDS_dom is a relevant proxy for Brazil risk, important components remain uncaptured by this set of factors, such as changes in the term premium, repricing of the monetary policy path, liquidity conditions, idiosyncratic news, and microstructure shocks.

**(ii) Negative trough in cumulative DI — 2020-08-03**

**DI5Y_change_cum = −802.0 bps**

Cumulative decomposition at this point:

| Accumulated Component | Value (bps) |
|---|---|
| Const_cum | 716,478 |
| Macro_cum | 277,586 |
| RiscoBR_cum | −36,9119 |
| Global_cum | −14,2008 |
| Residual_cum | −850,2938 |

*Table 5 – Cumulative decomposition up to 08/03/2020.*

**Interpretation:** The cumulative collapse in DI5Y through mid-2020 is largely concentrated in the residual. This is economically plausible: episodes of significant easing, term-premium repricing, and regime changes can occur with limited linear correspondence to risk proxies (CDS) and to daily changes in Focus expectations. In other words, the residual is not simply "error"; it indicates that the selected set of drivers does not intend (and cannot) capture the full dynamics of the cycle.

**(iii) End of sample — 2025-12-12**

DI5Y_change_cum = +7.5 bps

Cumulative decomposition at this point:

| Accumulated Component | Value (bps) |
|---|---|
| Const_cum | 1,409,811 |
| Macro_cum | 107,171 |
| RiscoBR_cum | −13,3540 |
| Global_cum | −130,8443 |
| Residual_cum | ≈ 0 |

*Table 6 – Cumulative decomposition up to 12/12/2025.*

The cumulative residual being close to zero at the end of the sample is an expected property when using OLS with an intercept on the same sample used for estimation: the sum of residuals tends toward zero, so the accumulated residual "closes" at the end. For economic interpretation, therefore, the focus should be on the path of the residual (rather than its final value) and on the balance across blocks over time.

## 5. DISCUSSION: WHAT THIS FRAMEWORK CAPTURES

### 5.1. The role of sovereign risk as a high-frequency driver

The results confirm that sovereign risk measures (especially the domestic component of CDS) are powerful in explaining daily movements in the intermediate/long end of the Brazilian yield curve. This is consistent with the literature documenting the sensitivity of sovereign spreads to global variables and to idiosyncratic components, particularly during stress periods (Jeanneret, 2018).

The decomposition into **CDS_glob** and **CDS_dom**, while parsimonious, provides a clear analytical narrative. When DI5Y moves due to predominantly external shocks—such as shifts in global risk appetite, a stronger dollar, or repricing of the U.S. curve—the model tends to attribute a meaningful share of the move to **CDS_glob**. Conversely, when the dynamics are essentially domestic—marked by fiscal uncertainty, political noise, or deterioration in local risk—**CDS_dom** becomes the main explanatory channel. This separation is particularly valuable in emerging economies, where similar shifts in the yield curve may originate from very different sources. In recent episodes, for instance, the combination of a "Trump Trade" with events that increased distrust regarding Brazil's fiscal balance could be quantified and analyzed separately, allowing one to measure the specific impact of each component on the curve—especially useful in environments of high complexity and rapid risk repricing.

### 5.2. The supervised macro factor: information compression, not causality

Using PLS addresses the practical need to reduce dimensionality without losing the connection to the target. Rather than manually choosing weights for IPCA, Selic, GDP, and fiscal variables across multiple horizons, PLS finds a vector of weights that maximizes covariance with ΔDI5Y. This produces an interpretable factor (with an anchored sign), but its supervised nature requires caution: the factor is "trained" to move with DI.

Consequently, the correct interpretation of the coefficient $\beta_M$ is: "the bps scale associated with one unit of the constructed macro factor." The statistical significance of $\beta_M$, although observed, should be interpreted as evidence that the macro signal is coherent with DI within the factor-construction process itself—not as strict causal evidence.

### 5.3. The residual as a completeness metric for the driver set

With **$R^2 \approx 0.2245$**, a substantial share (≈ **77.5%**) of the daily variance of ΔDI5Y remains in the residual. In an applied research design, this is informative in its own right: it indicates that the selected drivers (macro via expectations and risk via CDS) explain a material portion, but leave out important components such as:

- term premium and slope/curvature dynamics (term-structure factors);
- discrete monetary policy events (Copom) in announcement windows;
- domestic measures of financial conditions and liquidity;
- positioning and flows (technical factors);
- nonlinearities and regime shifts.

From the term-structure literature perspective, this is coherent: yield decompositions often separate short-rate expectations from the term premium and capture multiple factors (Abrahams et al., 2016; Joslin, Singleton, & Zhu, 2011). A three-driver model, by construction, is not intended to replicate the entire structure.

## 6. CONCLUSION

This article presented a complete method to decompose, in basis points, the daily changes in Brazil's 5-year DI Futures into three interpretable blocks: (i) **Macro/Central Bank**, summarized by a supervised factor built from changes in Focus expectations and a surprise measure; (ii) Brazil Risk, measured by the domestic component of CDS (the residual after removing the global component); and (iii) External Risk, measured by the global component of CDS explained by proxies for international financial conditions.

In the effective sample (2015-01-13 to 2025-12-12, 2,741 observations), the baseline model explains about 22.45% of the daily variance of ΔDI5Y. Within the explained portion, domestic risk is the dominant driver, global risk is complementary, and the supervised macro factor contributes on a smaller scale, albeit with statistical coherence. The cumulative decomposition produces a useful quantitative narrative—especially for historical interpretation and communication—while also making the residual explicit as a "thermometer" of what is left outside the chosen variable set.

The final message is twofold. On the one hand, it is possible to build a transparent and replicable decomposition of the curve (or, here, DI5Y) in bps using a relatively compact set of drivers. On the other hand, the complexity of the term structure in emerging markets imposes clear limits: cycles and regime changes, term premia, and discrete events generate movements that will not be captured by a linear specification with only a few factors. In future versions, rolling estimations, robust standard errors, and enriching the driver set are natural extensions to bring the framework closer to a stronger evidentiary standard and to the applied frontier.